\documentclass{article}
\usepackage{dcase2020_techrep,amsmath,graphicx,url,times,booktabs, tabularx}
\usepackage{tablefootnote}
\usepackage{amsmath,bm}
\usepackage{hyperref}
\usepackage{amssymb}

\title{ERANNs: Efficient Residual Audio Neural Networks for Audio Pattern Recognition}

\name{Sergey Verbitskiy$^{1, 3}$, \thanks{Corresponding author:
s.verbitskii@alumni.nsu.ru (Sergey Verbitskiy)}
       Vladimir Berikov$^{2}$,
       Viacheslav Vyshegorodtsev$^{3}$, 
       }
 
 \address{$^1$ Novosibirsk State University, Department of Mechanics and Mathematics, Novosibirsk, Russia, \\      
         $^2$ Sobolev Institute of Mathematics SB RAS, Data Analysis Laboratory, Novosibirsk, Russia, \\
         $^3$ Deepsound, Novosibirsk, Russia, \\
  }

\begin{document}

\ninept
\maketitle
\begin{sloppy}
\begin{abstract}
Audio pattern recognition (APR) is an important research topic and can be applied to several fields related to our lives. Therefore, accurate and efficient APR systems need to be developed as they are useful in real applications. In this paper, we propose a new convolutional neural network (CNN) architecture and a method for improving the inference speed of CNN-based systems for APR tasks. Moreover, using the proposed method, we can improve the performance of our systems, as confirmed in experiments conducted on four audio datasets. In addition, we investigate the impact of data augmentation techniques and transfer learning on the performance of our systems. Our best system achieves a mean average precision (mAP) of 0.450 on the AudioSet dataset. Although this value is less than that of the state-of-the-art system, the proposed system is 7.1x faster and 9.7x smaller. On the ESC-50, UrbanSound8K, and RAVDESS datasets, we obtain state-of-the-art results with accuracies of~0.961, 0.908, and 0.748, respectively. Our system for the ESC-50 dataset is 1.7x faster and 2.3x smaller than the previous best system. For the RAVDESS dataset, our system is 3.3x smaller than the previous best system. We name our systems ``Efficient Residual Audio Neural Networks''.

~
\end{abstract}

\section{Introduction}
Studies on audio pattern recognition (APR) have been gradually increasing, and APR systems can be used in several fields that are related to our everyday lives. The primary research subtopics of APR are environmental sound classification~\cite{10.1145/2733373.2806390, 10.1145/2647868.2655045}, sound event detection~\cite{Turpault2019_DCASE}, and audio tagging (predicting the presence or absence of sounds in audio signals)~\cite{7952261}. In general, we are surrounded by environmental sounds; therefore, there are several applications of APR systems, e.g., smart room monitoring~\cite{cite6} and video content highlight generation~\cite{5202537}. Moreover, over the past decade, several studies have focused on musical genre classification~\cite{1021072}. Therefore, APR systems can be used to analyze music preferences of users of media services to improve the recommendation process. Another application of APR systems is speech emotion classification~\cite{Livingstone2018TheRA}. In addition, APR systems can be used in the medical field, e.g., to classify respiratory diseases ~\cite{10.1007/978-981-10-7419-6_6}. In this study, we cover three APR subtasks, for which our systems are evaluated: audio tagging (AudioSet~\cite{7952261}), environmental sound classification (\hbox{ESC-50~\cite{10.1145/2733373.2806390}} and UrbanSound8K~\cite{10.1145/2647868.2655045}), and speech emotion classification (RAVDESS~\cite{Livingstone2018TheRA}).

The AudioSet dataset~\cite{7952261}, which is the largest dataset containing two million audio recordings with 527 classes, is used to train and evaluate modern APR systems. Recently, several researchers conducted studies~\cite{gong21b_interspeech, 9229505, 9414229, Guzhov2021ESResNeXtfbspLR} to obtain high performance on the aforementioned dataset. Despite a considerable amount of effort to make APR systems more accurate for AudioSet, the systems still have significantly lower performance compared with modern systems used for image recognition tasks on datasets that are of a similar size. For instance, the state-of-the-art model achieves a mean average precision (mAP) of 0.485 on AudioSet~\cite{gong21b_interspeech}; however, previous image classification systems achieve a top-1 accuracy of 0.9~\cite{Pham_2021_CVPR} on ImageNet~\cite{5206848}. Therefore, the APR field needs to be explored further to develop more accurate systems.

In addition to the performance of APR systems, we need to focus on the computational complexity of the systems. In recent years, several studies focused on techniques to reduce the computational complexity of neural networks for diverse tasks~\cite{pmlr-v97-tan19a}, including APR tasks~\cite{9414229}, as it is important to implement systems for low-powered devices and reduce the use of computing resources in data centers. However, the best system for AudioSet~\cite{gong21b_interspeech} has low efficiency, which makes it practically infeasible to use this system in real life. Thus, there is a need to develop APR systems, which are also compact and fast.

Considering the aforementioned issues, we suggest several techniques to improve the performance and efficiency of APR systems. First, we develop a new convolutional neural network (CNN) architecture with the widening factor~\cite{zagoruyko2017wide}, which is used to change the number of parameters and floating-point operations (FLOPs) of systems. Moreover, we propose a simple method to improve the inference efficiency of CNN-based systems for APR tasks. This method is based on an increase in the stride size (from 2 to 4) of few convolutional layers. Second, we describe and compare several data augmentation techniques for audio pattern recognition tasks, which are applied to improve the accuracy of systems on evaluation sets: the temporal cropping, SpecAugment~\cite{Park2019SpecAugmentAS}, modified mixup~\cite{tokozume2018learning}, and pitch shifting~\cite{810857}. To estimate the impact of diverse techniques on the performance and computational complexity of APR systems, we conduct experiments on four benchmark audio datasets for tagging and classification tasks:  AudioSet~\cite{7952261}, \hbox{ESC-50~\cite{10.1145/2733373.2806390}}, \hbox{UrbanSound8K~\cite{10.1145/2647868.2655045}}, and RAVDESS~\cite{Livingstone2018TheRA}. In addition, transfer learning is applied to \mbox{ESC-50}, UrbanSound8K, and RAVDESS: we transfer our systems, which are pre-trained on AudioSet. Thus, we examine the impact of transfer learning on the performance of systems on small APR datasets. Our APR systems are known as \hbox{``Efficient Residual Audio Neural Networks'' (ERANNs)}.

Owing to the design and application of various techniques, ERANNs can obtain high results on all the datasets. The proposed system achieves an mAP of 0.450 on the AudioSet dataset. Although this value is less than that of the state-of-the-art system (0.485)~\cite{gong21b_interspeech}, our system is 7.1x faster and has 9.7x fewer parameters. Moreover, the best system is pre-trained on ImageNet~\cite{5206848}; however, our system is trained from scratch. On the ESC-50, UrbanSound8K, and RAVDESS datasets, we obtain state-of-the-art accuracies of~0.961, 0.908, and 0.748, respectively. Previous state-of-the-art systems achieve accuracies of 0.956~\cite{gong21b_interspeech}, 0.891~\cite{Guzhov2021ESResNeXtfbspLR}, and 0.721~\cite{9229505}, respectively. Our system for ESC-50 is 1.7x faster and 2.3x smaller than the previous best system. For RAVDESS, our system is 3.3x smaller than the previous best system.

The remainder of this paper is organized as follows. Section~\ref{section2} describes previous methods used for APR tasks. Section~\ref{section3} presents the feature extraction process, data augmentation techniques, and the proposed CNN architecture. Section~\ref{section4} provides detailed experimental results and a comparison of our APR systems with previous systems. Section~\ref{section5} concludes this paper.

\section{Related Works}
\label{section2}

Traditional APR systems comprised classical generative or discriminative models, e.g., Gaussian mixture models (GMMs)~\cite{Vuegen2013ANMA}, and used time-frequency representations, e.g., the log mel spectrogram or mel-frequency cepstral coefficients (MFCCs), as input. However, later, methods with neural networks, in particular with CNNs~\cite{tokozume2018learning, 9229505,9414229, 9413035, Guzhov2021ESResNeXtfbspLR}, significantly outperformed conservative machine learning methods for APR tasks. Moreover, CNN-based systems outperformed recurrent neural network RNN-based systems, e.g., for sound event detection tasks~\cite{rnn_comparison}. Nowadays, \hbox{CNN-based} systems are widely used for APR tasks.

Systems with transformers are gaining popularity for APR tasks. A transformer-based system achieves the state-of-the-art performance with an mAP of~0.485~\cite{gong21b_interspeech} on the AudioSet dataset. However, this system has high computational complexity (526.6 million parameters), which makes it difficult to apply this system in real life.

There are two main approaches for CNN-based systems. The first approach states to use models with 2D convolutional layers and time--frequency representations, e.g., the log mel spectrogram, as input to the first 2D convolutional layer~\cite{9229505, 9413035}. The second approach defines end-to-end systems~\cite{9229505, 9414229, ABDOLI2019252, tokozume2018learning}, where the raw audio signal is used as input. End-to-end systems, as a rule, comprise two parts~\cite{9229505, 9414229, tokozume2018learning}: the first part includes 1D convolutional layers, and the second part includes 2D convolutional layers. The first part is applied to extract 2D features, which replace \hbox{time--frequency} representations. Despite the learnable extraction of \hbox{time--frequency} features in end-to-end systems, systems with the log mel spectrogram perform better on APR tasks~\cite{9229505, 9414229}. Moreover, there are end-to-end systems, which contain only 1D convolutional layers~\cite{ABDOLI2019252}. In this article, we use the first approach and the log mel spectrogram as input to the first 2D convolutional layer. We compare the aforementioned approaches below.

Residual neural networks (ResNets)~\cite{7780459} have shortcut connections among convolutional layers. Shortcut connections help partially avoid the vanishing gradient problem. Over the past years, the application of shortcut connections in CNN architectures became a widespread technique in the area of machine learning, including APR~\cite{9229505, 9413035, Guzhov2021ESResNeXtfbspLR}. WideResNets~\cite{zagoruyko2017wide} have an additional hyperparameter---the widening factor for the width of convolutional layers---to change the computational complexity of models. With optimal values of the widening factor, WideResNets have better performance and fewer parameters than original ResNets. The widening factor was introduced in several CNN-based systems for simple scaling, for instance, for image recognition tasks~\cite{pmlr-v97-tan19a} (EfficientNet) and APR tasks~\cite{9414229} (AemNet-DW). We also use shortcut connections and the widening factor in our CNN architecture. 

Transfer learning is a widespread technique in computer vision tasks. A model is trained for tasks with a large dataset and transferred to similar tasks with smaller datasets: all parameters of the model for the new task are initialized from the pre-trained model, except parameters of few last layers. For APR tasks, transfer learning was used by other researchers~\cite{gong21b_interspeech, 9229505, 9414229, Guzhov2021ESResNeXtfbspLR} (pre-training on AudioSet) that significantly improved accuracy on APR tasks with small datasets. In this article, we use fine-tuning as the transfer learning strategy (we optimize all parameters of transferred models). Models pre-trained on the AudioSet dataset are transferred to the other three APR tasks with small datasets: ESC-50, \hbox{UrbanSound8K}, and RAVDESS.

\section{Proposed APR System}
\label{section3}
This section describes the feature extraction process, data augmentation techniques, and the proposed CNN architecture.

\subsection{Feature Extraction}
\label{section31}
For each experiment, we use the log mel spectrogram as input to the first 2D convolutional layer of our models.

We use a sampling rate ($sr$) of~44.1 kHz for all audio signals. The short-time Fourier transform (STFT) with the Hann window of size 1380 ($\approx$~31~ms) and the hop size of $H = 345$ ($\approx$~8~ms) is applied to extract spectrograms. The number of time frames $T_s$ is calculated as
\begin{equation}
\begin{gathered}
\label{eq:1}
T_s = \Bigg[ \frac{sr \cdot t}{H} \Bigg] + 1 = \Bigg[ \frac{44100 \cdot t}{345} \Bigg] + 1,
\end{gathered}
\end{equation}
where $t$ is the duration of an audio signal (in seconds). Integer $t \leq 5$ yields a value of $T_s = 128\cdot t$. For integer $t>5$, we pad the audio signal with zeros to obtain $T_s = 128\cdot t$. If $t$ is not an integer, we pad the audio signal with zeros to gain integer duration. Thus, \hbox{$T_s \, / \, 2^j \in \mathbb{N}$} for $j \leq 7$. This is a convenient way to avoid size mismatch in shortcut connections of models with different sets of stride sizes of convolutional layers for input audio signals with various lengths (in our models, the maximum product of stride sizes for the temporal dimension is 128). This substantiates the selection of $H = 345$. The size of the window is selected to ensure an overlap of 0.75.

We adopt the number of mel bins $M = 128$. We suppose that $M = 128$ is the best choice as a trade-off between computational complexity and performance of models. In~\cite{9229505}, the model with $M = 128$ performed better than the model with $M=64$ on the AudioSet dataset. Large values of $M$ can significantly decrease the efficiency of models.

Moreover, we determine the lower cut-off frequency \mbox{$f_{min} = 50$} Hz to remove low-frequency noise and the upper cut-off frequency \mbox{$f_{max} = \text{14,000}$ }Hz to remove the aliasing effect. 

Thus, the log mel spectrogram of each audio recording is represented as a 2D tensor with a shape of $128 \times T_s$.

\subsection{Data Augmentation Techniques}
\label{section32}
We use several data augmentation techniques to prevent models from overfitting during training and improve the performance of models on evaluation sets. All data augmentation techniques are applied during training on mini-batches of audio signals generated from original training sets but not before training to obtain more extensive training sets.

The first technique is the temporal cropping. We use \mbox{$t_c$-second} sections of audio recordings during training of our models. Sections are cut from random places. While evaluating models, full audio recordings without the temporal cropping are used to obtain final predictions. 

In our experiments, we use SpecAgment~\cite{Park2019SpecAugmentAS} for frequency and time masking on spectrograms of training audio clips. This technique was used in previous studies for APR tasks~\cite{9229505, gong21b_interspeech}. We use SpecAugment with two time masks with a maximum length of $8\cdot t_c$ frames and two frequency masks with maximum length of 16 bins.
	
Moreover, we apply modified mixup~\cite{tokozume2018learning}, which considers the sound pressure level of two mixed audio signals. We use mixup with $\alpha=1.0$ in all the experiments. In~\cite{tokozume2018learning}, models trained using modified mixup had better performance than models trained with standard mixup on the ESC-50 and UrbanSound8K datasets. Moreover, we show that modified mixup is better for the AudioSet dataset below. In previous studies, for AudioSet, only the standard mixup was used~\cite{9229505, gong21b_interspeech, 9414229}.

Pitch shifting~\cite{810857} is used for RAVDESS during training because this data augmentation technique is effective for speech, as described in the following. This data augmentation technique is applied with a probability of 0.5 for each audio signal.

\subsection{ERANNs Architecture}
Each tensor between convolutional blocks has a shape of \mbox{$F_{i}\times T_i \times C_{i}$}, where $F_{i}$ is \hbox{``the i-th  frequency size''}, $T_i$ is ``the i-th temporal size'' and $C_{i}$ is the number of channels. All input tensors after feature extraction have a shape of $F_0\times T_{0} \times C_0$, where $F_0 = M = 128$, $T_0 = T_s$, and $C_0 = 1$.

We modify sizes of kernel, stride, and padding for the temporal dimension in few convolutional layers to make models faster. We introduce ''the decreasing temporal size parameter'' $s_m$ into our architecture, and for $s_m>0$, temporal sizes $T_i$ are reduced stronger based on increased stride sizes of convolutional layers. Thus, increasing the value of $s_m$, we reduce FLOPs of our models because sizes of tensors between convolutional layers have fewer sizes. 

The proposed architecture is presented in Table~\ref{tab:tablearch}.

\vspace{-1.5em}
\begin{table}[h]
  \caption{Proposed ERANNs architecture}
  \centering
  \begin{tabular}{ l c c }
  \hline
    \textbf{Stage name}& \textbf{Output size} & 
                                         \textbf{Layers} \\
    \hline
    Extraction      & $128 \times T_0 \times 1 $ & \begin{tabular}[c]{@{}c@{}} Feature extractor\\ BatchNorm \end{tabular} \\ \hline
    Stage 0     & $128 \times T_0 \times 8 \cdot W$& \begin{tabular}[c]{@{}l@{}} ARB(1, 1, $8\cdot W) \times 4$ \end{tabular}              \\ \hline
    Stage 1  & $64 \times  T_1 \times 16 \cdot W $ & 
\begin{tabular}[c]{@{}l@{}} ARB(2, $s_1$, $16 \cdot W) \times 1$\\ ARB(1, 1, $16\cdot W) \times 3$
\end{tabular} \\ \hline
Stage 2 & $32 \times  T_2 \times 32\cdot W $&  
\begin{tabular}[c]{@{}l@{}} ARB(2, $s_2$, $32 \cdot W) \times 1$ \\ ARB(1, 1, $32\cdot W) \times 3$
\end{tabular} \\ \hline
Stage 3& $16 \times T_3 \times 64\cdot W $&  
\begin{tabular}[c]{@{}l@{}} ARB(2, $s_3$, $64 \cdot W) \times 1$ \\ ARB(1, 1, $64 \cdot W) \times 3$
\end{tabular} \\ \hline
Stage 4& $8 \times T_4  \times 128\cdot W $ &  
\begin{tabular}[c]{@{}l@{}} ARB(2, 2, $128 \cdot W) \times 1$ \\ ARB(1, 1, $128\cdot  W) \times 3$
\end{tabular} \\ \hline
 & $1 \times 1 \times 128\cdot W $ &  Global pooling  \\
Stage 5& $128\cdot W$ & FC1, Leaky ReLU \\  
&$N$ & FC2, sigmoid / softmax \\
    \hline
  \end{tabular}
  \label{tab:tablearch}
  
\end{table}

The input of our ERANNs is a mini-batch with audio signals. Data augmentation for training data is implemented for each mini-batch before the feature extractor. SpecAugment is applied after the extraction of spectrograms.

In the table, $W$ is the widening factor and $s_{i}$ are stride sizes, which depend on the decreasing temporal size parameter $s_m$ that can have four values:
\begin{equation}
s_m \in \{0,...,3\}.
\end{equation}

Stride sizes $s_i$ for $i = 1, ..., 3$ for convolutional blocks \mbox{ARB($2$, $s_i$, $c$)} can be calculated as
\begin{equation}
\begin{gathered}
s_i = 
 \begin{cases}
   2 & i > s_m\\
   4 & i \leq s_m.
 \end{cases}  
 \end{gathered}
\end{equation}
Temporal sizes $T_i$ for $i = 0, ..., 3$ are calculated as
\begin{equation}
T_0 = T_s,\;\;\;\; T_i = \frac{T_0}{\prod_{j=1}^{i}{s_j}}, \; i\geq 1\\.
\end{equation}

ARB($x$, $y$, $c)$ is ``the Audio Residual Block'', where $x$ is the stride size for the frequency dimension, $y$ is the stride size for the temporal dimension, and $c$ is the number of output channels. Each ARB contains two or three 2D convolutional layers, two batch normalization layers, and two activation functions. The Audio Residual Block is the modified version of the basic block of \mbox{ResNet-V2}~\cite{10.1007/978-3-319-46493-0_38}. The ARB architecture is described in Fig.~\ref{fig:ARB} (Table~\ref{tab:tableARB}). 
\vspace{-1.5em}
\begin{table}[h!]
  \caption{Proposed ARB(x,y,c) architecture}
  \centering
  \begin{tabular}{ l c c c }
  \hline
\textbf{}&  \textbf{Kernel}& \textbf{Stride} & 
                                         \textbf{Padding} \\
    \hline
BatchNorm$_1$ & $-$ &  $-$ & $-$  \\ 
Leaky ReLU     & $-$ &  $-$ & $-$  \\ 
Conv2D$_1$    & $K_1(x)\times K_1(y)$ &  $x\times y$ & $1\times 1$  \\ 
BatchNorm$_2$      & $-$ &  $-$ & $-$  \\ 
Leaky ReLU      & $-$ &  $-$ & $-$  \\ 
Conv2D$_2$      & $K_2(x)\times K_2(y)$ &  $1\times 1$ & $P(x)\times P(y)$  \\ 
Conv2D$_{Res}$      & $1\times 1$ &  $x\times y$ & $0\times  0$.  \\ 
    \hline
  \end{tabular}
  
  \label{tab:tableARB}
   \vspace{-2mm} 
\end{table}

\begin{figure}[h!]
  \centering
  \centerline{\includegraphics[width=0.85\columnwidth]{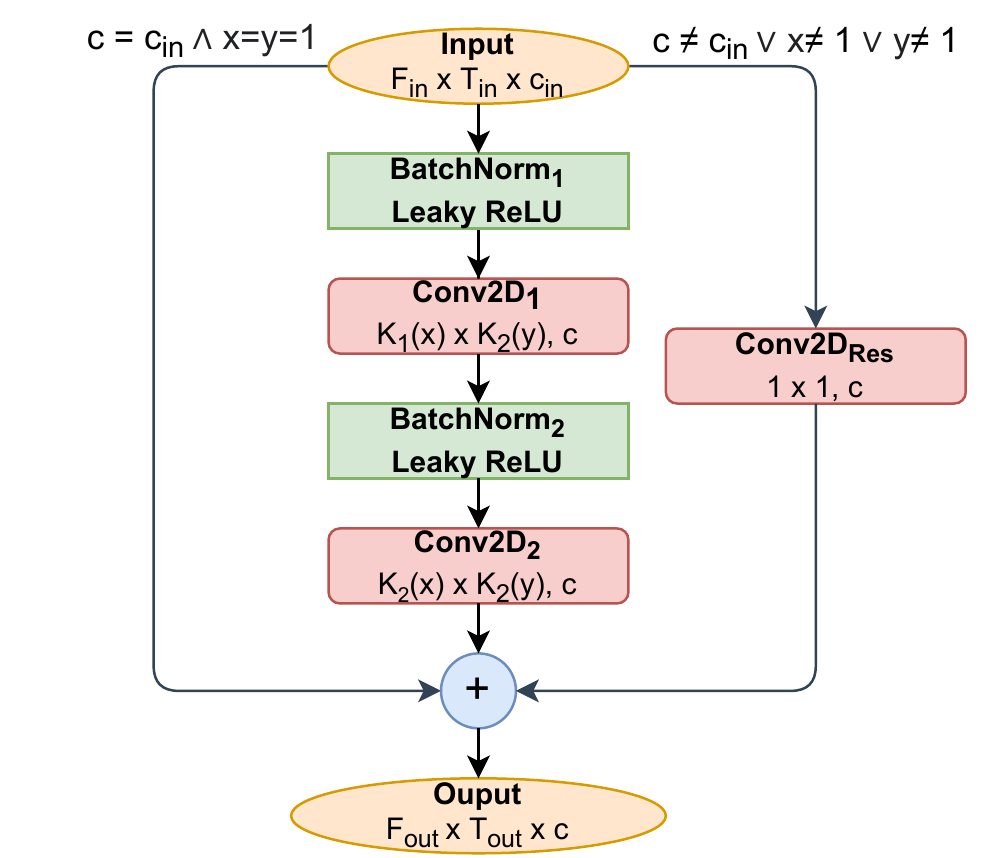}}
  \caption{Scheme of ARB(x,y,c)}
  \label{fig:ARB}
\end{figure}

Functions $K_{1}(z)$, $K_{2}(z)$, and $P(z)$ for ARBs, where $z$ is the stride size, are defined as
\begin{equation}
(\,K_1(z), \, K_2(z), \, P(z) \,) = 
 \begin{cases}
   (\,3,\, 3,\, 1\,), & \text{if } \, z=1 \lor z=2\\
   (\,6, \,5, \,2\,), & \text{if } \, z=4.
 \end{cases}  
\end{equation}

We do not use dimension reduction as in bottleneck blocks because \cite {9229505} stated that ResNet38 with basic blocks performs better than ResNet54 with bottleneck blocks on the AudioSet dataset.

The first ARB does not have BatchNorm and Leaky ReLU at the start. The batch normalization layer after the feature extractor is used as a replacement for data standardization (over the frequency dimension). 

For global pooling, we use a sum of average and max pooling as in~\cite{9229505} to combine their advantages. After global pooling, we use two fully connected layers, FC1 and FC2. We apply softmax for sound classification tasks and sigmoid for audio tagging tasks at the end to obtain predictions. 

Leaky ReLU with parameter 0.01 is applied as a replacement for ReLU because the use of ReLU can lead to the dying ReLU problem~\cite{Lu_2020}.

The proposed architecture of ERANNs does not depend on the duration of input audio signals caused by the presence of global pooling before the last fully connected layers. For instance, the system with fixed values of $W$, $s_m$, and $N$ can be trained with 8-second audio signals and applied to audio signals of any length. Audio signals are padded with zeros, as described in Section~\ref{section31}, where the number of zeros does not exceed 44,100 (1 second). FLOPs and the inference speed of the system with fixed values of hyperparameters depend linearly on the length of input audio signals, excluding the computational cost of the last fully connected layers.

\section{Experiments and Results}
\label{section4}

In this section, we demonstrate the comparison of the performance and computational complexity of ERANNs with various values of hyperparameters $W$ and $s_m$ for four datasets: AudioSet~\cite{7952261}, ESC-50~\cite{10.1145/2733373.2806390}, UrbanSound8K~\cite{10.1145/2647868.2655045}, and RAVDESS~\cite{Livingstone2018TheRA}. Moreover, we compare the performance of models for AudioSet, which are trained with various data augmentation techniques. Models pre-trained on AudioSet are transferred to the other three datasets to obtain higher performance. Finally, we compare our ERANNs with previous state-of-the-art APR systems for all the datasets. We show that the proposed techniques are effective for APR tasks.

Each system with fixed values of hyperparameters is abbreviated as ERANN-$s_{m}$-$W$, where $s_m$ is the decreasing temporal size parameter and $W$ is the widening factor.

\subsection{Experimental Setup}

Models parameters are optimized by minimizing the categorical cross-entropy loss with the Adam optimizer~\cite{DBLP:journals/corr/KingmaB14} and a mini-batch size of~32. We use the one-cycle learning rate policy~\cite{smith2018superconvergence} for the AudioSet dataset with a maximum learning rate of~0.001. For \hbox{ESC-50}, UrbanSound8K, and RAVDESS, we use a constant learning rate of~0.0002 for training from scratch and a constant learning rate of~0.0001 for fine-tuning. 

In addition, final models for evaluation sets are obtained using an exponential moving average (EMA) with a decay rate of~0.999 to increase the sustainability of models for evaluation sets and avoid the considerable influence of last training iterations. Thus, the value of the parameter $\tilde{\theta}_j$ of the model for the evaluation set for the $j-$th iteration is calculated as
\begin{equation}
\begin{gathered}
\tilde{\theta}_j = 
 \begin{cases}
   \theta_0 , & \text{if } \, j=0\\
   \beta \, \tilde{\theta}_{j-1} + (1-\beta) \, \theta_j ,& \text{otherwise},
 \end{cases} 
\end{gathered}
\end{equation}
where $\theta_j$ is the value of the corresponding learnable parameter for the $j-$th training iteration and $\beta=0.999$ is the decay rate.

\subsection{AudioSet}
\setlength{\parindent}{3.0ex}
AudioSet~\cite{7952261} is a large-scale audio dataset. The dataset includes over 2 million audio recordings (2,085,999 audio recordings in the training set and 20,371 audio recordings in the evaluation set) with 527 sound classes. Audio recordings are extracted from YouTube videos, and the duration of most audio clips is 10 seconds. 

We successfully downloaded 87.4\% of audio clips of the training set and 88.2\% of audio clips of the evaluation set. All audio recordings are converted into the monophonic format at a sampling rate of~44.1 kHz.

The AudioSet dataset is a multilabel dataset (tagging task). Therefore, sigmoid is used to obtain predictions of models.

We use a balanced sampling strategy for the AudioSet dataset during training~\cite{9229505} because this dataset is highly unbalanced. As described in~\cite{9229505}, the balanced sampling strategy significantly improves the performance of models.

Our models are trained for 500,000 iterations with the full training set without cross-validation. Final models are obtained using early stopping (models are evaluated on the evaluation set at an interval of 5000 iterations). All experiments are repeated three times with different random seeds to obtain final results (the mean of three results). To evaluate and compare APR systems on the evaluation set, we use the mean average precision (mAP) as the main evaluation metric. 

\subsubsection{Ablation Studies}

We conduct an ablation study to examine the impact of data augmentation techniques and various values of hyperparameters on the performance and computational complexity of our models.

\vspace{3mm}

\noindent\textbf{Impact of Data Augmentation Techniques:} We compare the performance of models trained with various data augmentation techniques. We conduct experiments with different types of mixup for \mbox{ERANN-2-5} ($t_c = 8$): 1---training without mixup, 2---standard mixup on the waveform, 3---modified mixup on the waveform, and 4---modified mixup on the log mel spectrogram. We also show advantages of using SpecAugment for \mbox{ERANN-2-5} ($t_c = 8$) and the temporal cropping for \mbox{ERANN-1-6} ($t_c \in \{4, 8, 10\}$) during training. The results of all the experiments are listed in Table~\ref{tab:tableaug}.

\vspace{-1.5em}
\begin{table}[h!]
  \caption{Comparison of data augmentation techniques for AudioSet}
  
  \centering
  \begin{tabular}{lcccc}
  \hline
\textbf{}&  \textbf{$\bf{t_c}$}& \textbf{SpecAugment} & 
                                         \textbf{mixup type}  & 
                                         \textbf{mAP}\\
    \hline
ERANN-2-5      & 8 &  $\checkmark$ & 1 & 0.430\\ 
ERANN-2-5     & 8 &  $\checkmark$ & 2 & 0.443 \\ 
ERANN-2-5    & 8 &  $\checkmark$ & 3 &\textbf{0.446}\\  
ERANN-2-5      & 8 &  $\checkmark$ & 4 & 0.439\\ 
\hline
ERANN-2-5     & 8 & $\times$ & 3  & 0.441 \\ 
\hline
ERANN-1-6      & 4 &  $\checkmark$ & 3  & 0.417\\  
ERANN-1-6      & 8 &  $\checkmark$ & 3  & \textbf{0.450}\\  
ERANN-1-6      & 10 &  $\checkmark$ & 3 &0.440 \\\hline
  \end{tabular}
  \label{tab:tableaug}
\end{table}

The model trained using the modified mixup on the waveform and SpecAugment has the best performance. It can be concluded that $t_c = 8$ is the best value in terms of the duration of cropped sections of training audio recordings among the values of 4, 8, \hbox{and 10}.

\vspace{3mm}

\noindent \textbf{Impact of Hyperparameters:} We compare the performance, the number of parameters, and the inference speed of ERANNs with various values of two hyperparameters. Models are compared with three values of $W$ with $s_{m}=4$ and four values of $s_{m}$ with $W=6$. We stop at $W=6$, so that the model is not too large, and stop at $W=4$, so the model is not too simple. To evaluate the inference speed of models, we calculate the number of resulting predictions for 10-second audio recordings per second. We use \hbox{NVIDIA Tesla V100} and compare the inference speed of models for mini-batch sizes of $B=1$ and $B=32$. Finally, we compare our APR systems with AST systems~\cite{gong21b_interspeech}, which have state-of-the-art performance on the AudioSet dataset. The results are demonstrated in Table~\ref{tab:tableeff}.

\vspace{-1.5em}
\begin{table}[h!]
  \caption{Comparison of the computational complexity and the performance of ERANNs on AudioSet (OOM is Out Of Memory)}
  \centering
  \begin{tabular}{lcccc}
  \hline
 & \textbf{Params} & \multicolumn{2}{c}{\textbf{Audios/sec}}  & \textbf{mAP}\\
  &  \textbf{\footnotesize{$\bf{\times \, 10^6}$}} & \textbf{\footnotesize{B=1}}  &\textbf{\footnotesize{B=32}} & \\ \hline 
ERANN-2-4    & 24.5 & 137 &  246  & 0.430 \\  
ERANN-2-5   & 38.2 & 96 & 149  & 0.446  \\ 
ERANN-0-6            & 54.4 & 53 &  74  & 0.447   \\
ERANN-1-6            & 54.5 &  71 & 108   & \textbf{0.450}\\
ERANN-2-6            & 54.9 & 82  &  124  & 0.448     \\
ERANN-3-6            & 56.5 & 79 &   128& 0.436      \\ \hline
AST~\cite{gong21b_interspeech}    & 88.1 &  49 &  \small{OOM}  &  0.459\\
AST (Ensemble-M)~\cite{gong21b_interspeech}  & 526.6 & 10  & \small{OOM}   & 0.485 \\ \hline
  \end{tabular}
  \label{tab:tableeff}
\end{table}

Increasing $s_m$ reduces FLOPs and increases the inference speed of models because tensors between convolutional layers have fewer sizes. As listed in Table~\ref{tab:tableeff}, \mbox{ERANN-1-6} and \mbox{ERANN-2-6} are 1.68x and 1.46x faster (for $B=32$) than \mbox{ERANN-0-6}, respectively. Moreover, \mbox{ERANN-1-6} and \mbox{ERANN-2-6} have better performance than ERANN-0-6.

The system AST~\cite{gong21b_interspeech} has better performance than \mbox{ERANN-1-6}; however, our system has 1.62x fewer parameters and is 1.45x faster. \mbox{ERANN-1-6} is significantly more efficient than the state-of-the-art system AST (\mbox{Ensemble-M})~\cite{gong21b_interspeech}: \mbox{ERANN-1-6} is 7.10x faster (for $B=1$) and 9.66x smaller.

As can be seen from Table~\ref{tab:tableeff}, using the proposed method, we can not only increase the inference speed of our APR systems, but also save and even improve their performance.

\subsubsection{Results}
The comparison of our ERANNs with previous APR systems is shown in Table~\ref{tab:tableresults}. Moreover, we compare end-to-end systems (''e2e'' in the table) and systems that use 2D time-frequency representations. 

\vspace{-1.5em}
 \begin{table}[th]
  \caption{Comparison of APR systems for AudioSet}
  \centering
  \begin{tabular}{lccc}
  \hline
& \textbf{e2e} & \textbf{Params} & \textbf{mAP}\\
  & &  \textbf{\footnotesize{$\bf{\times \, 10^6}$}} & \\ 
    \hline
ESResNeXt~\cite{Guzhov2021ESResNeXtfbspLR}& $\times$ & 31.1 & 0.282 \\
AemNet-DW (WM=1)~\cite{9414229} & $\checkmark$ & 1.2 & 0.329 \\
AemNet-DW (WM=2)~\cite{9414229} & $\checkmark$ & 3.0 & 0.340 \\
Wavegram-CNN~\cite{9229505} & $\checkmark$ & 81.0 &  0.389\\
CNN14~\cite{9229505} & $\times$ & 80.7 & 0.431\\
Wavegram-Logmel-CNN~\cite{9229505} & $\times$ & 81.1 & 0.439\\
AST~\cite{gong21b_interspeech}  & $\times$ & 88.1 & 0.459 \\
AST (Ensemble-M)~\cite{gong21b_interspeech} & $\times$ & 526.6 & \textbf{0.485} \\
\hline

ERANN-2-5  &$\times$&38.2 & 0.446\\
ERANN-1-6  &$\times$& 54.5 & \textbf{0.450}\\ \hline
 \end{tabular}
  
  \label{tab:tableresults}
 
\end{table}

Our system \mbox{ERANN-1-6} does not achieve the state-of-the-art results on the AudioSet dataset; however, our system is significantly more efficient than the state-of-the-art system, AST~\cite{gong21b_interspeech}. Moreover, AST was pre-trained on ImageNet~\cite{5206848}; our system is trained from scratch. 

The proposed APR system \mbox{ERANN-2-5} with an mAP of~0.446 is 2.11x smaller than CNN14 and Wavegram-Logmel-CNN~\cite{9229505} and has better performance. Moreover, our systems outperform other previous systems~\cite{Guzhov2021ESResNeXtfbspLR, 9414229} by a large margin.

We can see that end-to-end systems AemNet-DW~\cite{9414229} are significantly smaller than previous state-of-the-art systems, and our systems, which use 2D time-frequency representation as input. On the other hand, systems AemNet-DW are significantly less accurate.

\subsection{ESC-50}
\setlength{\parindent}{3.0ex}
ESC-50 (Environmental Sound Classification)~\cite{10.1145/2733373.2806390} is the dataset with environmental sounds comprising~50 sound classes and 2000 \mbox{5-second} audio recordings. This dataset is balanced with 40 audio recordings per sound class.

For each experiment, we repeat 5-fold cross-validation (official division into folds suggested by the authors of ESC-50) three times and calculate the average accuracy score (15 trainings for each model). All models are trained with four folds for 20,000 iterations from scratch and 2000 iterations using fine-tuning. We also use early stopping to obtain final models.

\subsubsection{Ablation Studies}
For ESC-50, we also determine the impact of various values of hyperparameters and pre-training on the AudioSet dataset on the performance and computational complexity of models.

\vspace{3mm}

\noindent\textbf{Impact of Hyperparameters and AudioSet Pre-training:} We compare the computational complexity and performance of models with various values of hyperparameters and the performance of models trained from scratch and fine-tuned models. For all the experiments, we adopt $t_c=4$ and use SpecAugment and modified mixup on the waveform during training. The comparison is shown in Table~\ref{tab:tableeffesc}.

\vspace{-1.5em}
\begin{table}[h!]
  \caption{Comparison of the computational complexity and the performance of ERANNs on ESC-50}
  \centering
  \scalebox{1.00}{
  \begin{tabular}{lcccc}
  \hline
& \textbf{Params} &\textbf{Audios/sec} & \multicolumn{2}{c}{\textbf{Accuracy}}\\
           &  \textbf{\footnotesize{$\bf{\times \, 10^6}$}}  &  \textbf{\footnotesize{B=32}}  & \textbf{\footnotesize{Scratch}} & \textbf{\footnotesize{Fine-tune}} \\
        \hline
ERANN-0-3            & 13.5 &   421 &0.882 & --  \\
ERANN-1-3            & 13.6 &  583    &\textbf{0.892} & --  \\ 
ERANN-0-4            & 24.0 &   300 & 0.892 & --  \\ 
ERANN-1-4            & 24.1 &   428  &0.891 & --   \\  
ERANN-2-4            & 24.3 &    482   & 0.880 & 0.949 \\ 
ERANN-2-5            & 37.9 &  294 & 0.879 & \textbf{0.961} \\ 
ERANN-0-6            & 54.1 &   150  & -- & 0.953    \\
ERANN-1-6            & 54.2 &   218 &  -- & 0.954   \\
ERANN-2-6            & 54.6 &   245  & -- & 0.959\\ \hline
AST~\cite{gong21b_interspeech}          & 87.3 &  176  & -- & 0.956\\ \hline
  \end{tabular}}
  \label{tab:tableeffesc}
\end{table}

ERANNs with $W\geq 5$ are large for the small dataset to train from scratch. High values of $W$ increase the fast overfitting risk. \mbox{ERANN-1-3} has higher accuracy than \mbox{ERANN-0-3} by 0.01 and is 1.38x faster. The system \mbox{ERANN-1-3} has 1.79x fewer parameters and is 1.94x faster than the system \mbox{ERANN-1-4}; however, \mbox{ERANN-1-3} achieves similar accuracy. \mbox{ERANN-2-6} trained using transfer learning is 1.63x faster than \mbox{ERANN-0-6}, while \mbox{ERANN-2-6} has better performance. Our best APR system for ESC-50 is fine-tuned \mbox{ERANN-2-5}. It has better performance, 2.30x fewer parameters, and is 1.67x faster than the previous best system AST~\cite{gong21b_interspeech}.

The use of pre-trained models significantly improve performance, while models are trained 10x faster: 20,000 iterations for training from scratch and 2000 for training using fine-tuning.

For ESC-50, the proposed method helps greatly reduce the computational complexity as well as increase the performance of our APR systems.

\subsubsection{Results}
The comparative analysis of ERANNs with previous APR systems for the ESC-50 dataset is shown in Table~\ref{tab:escfine}.

\vspace{-1.5em}
\begin{table}[th]
  \caption{Comparison of APR systems for ESC-50}
  \centering
   \scalebox{1.00}{
  \begin{tabular}{lcccc}
\hline
&\textbf{e2e} & \textbf{Params} &\multicolumn{2}{c}{\textbf{Accuracy}} \\
    
&&\textbf{\footnotesize{$\bf{\times \, 10^6}$}}&\textbf{\footnotesize{Scratch}}  &  \textbf{\footnotesize{Fine-tune}} \\ \hline
EnvNet-v2~\cite{tokozume2018learning} & $\checkmark$ & 101.3 & 0.818 & --\\
ESResNet~\cite{9413035} & $\times$ & 30.6 & 0.832 & 0.915\\ 
AemNet-DW~\cite{9414229} & $\checkmark$ & 0.9 & 0.749 & 0.923\\
CNN14~\cite{9229505} & $\times$& 79.8 & 0.833 &  0.947 \\
ESResNeXt~\cite{Guzhov2021ESResNeXtfbspLR}& $\times$ & 30.1 & -- & 0.952 \\
AST~\cite{gong21b_interspeech}&$\times$& 87.3& 0.887 & 0.956 \\
 \hline
ERANN-1-3 &$\times$&13.6& \textbf{0.892} & --\\
ERANN-2-5 &$\times$&37.9& 0.879 & \textbf{0.961}\\  
 \hline
  \end{tabular}}
  \label{tab:escfine}
\end{table}

Our system \mbox{ERANN-2-5} trained using transfer learning achieves an accuracy of~0.961, outperforming the previous state-of-the-art system AST~\cite{gong21b_interspeech}, which has an accuracy of~0.956.

Moreover, we can observe a significant difference between the performance of end-to-end systems and systems that use 2D time-frequency representations. At the same time, the end-to-end system \mbox{EnvNet-v2}~\cite{tokozume2018learning} has a large number of parameters; however, \mbox{AemNet-DW}~\cite{9414229} is very efficient compared with other systems listed in the table.

\subsection{UrbanSound8K}

The UrbanSound8K dataset includes 8,732 audio clips with 10 classes of urban sounds. Audio signals have various durations, which do not exceed 4 seconds. We convert all the recordings into the monophonic format at a sample rate of~44.1~kHz.

This dataset is divided into 10 folds by its authors that we applied to train and evaluate our models. All models are trained with 20,000 iterations from scratch and 5000 iterations using fine-tuning. All experiments are repeated three times.

\subsubsection{Results}

For UrbanSound8K, we use modified mixup on the waveform and SpecAugment as data augmentation techniques. We do not employ the temporal cropping because the duration of audio signals significantly varies (from 0.05 to 4 seconds).

In Table~\ref{tab:urban} we compare our ERANNs with various values of hyperparameters and previous systems for UrbanSound8K. We compare our systems with the previous systems, which were evaluated using 10-fold cross-validation without custom generation of segments as another way of evaluating systems can lead to unreasonably high performance~\cite{9413035}.

\vspace{-1.5em}
\begin{table}[th]
  \caption{Comparison of APR systems for UrbanSound8K}
  \centering
   \scalebox{1.00}{
  \begin{tabular}{ lcccc}
\hline
& \textbf{e2e}  & \textbf{Params} &  \multicolumn{2}{c}{\textbf{Accuracy}} \\
& &\textbf{\footnotesize{$\bf{\times \, 10^6}$}} &\textbf{\footnotesize{Scratch}} & \textbf{\footnotesize{Fine-tune}} \\ \hline
EnvNet-v2~\cite{tokozume2018learning} & $\checkmark$ & 101.1 & 0.783 & --\\
AemNet-DW~\cite{9414229} & $\checkmark$ & 0.9 & 0.763 & 0.835\\
ESResNet~\cite{9413035} & $\times$ & 30.6 & 0.828 & 0.854\\ 
ESResNeXt~\cite{Guzhov2021ESResNeXtfbspLR} & $\times$ & 30.0 & -- & 0.891\\ 
 \hline
ERANN-0-3 & $\times$ & 13.5 & 0.823 & --\\
ERANN-1-3 & $\times$ & 13.5 & 0.829 & --\\
ERANN-0-4 &$\times$ & 24.0 & 0.826  & -- \\
ERANN-1-4 &$\times$& 24.1 & \textbf{0.835} & --\\
ERANN-2-4 &$\times$ & 24.2 & 0.826 & 0.897 \\
ERANN-2-5 & $\times$& 37.9 & 0.813 & 0.903\\
ERANN-0-6 &$\times$& 54.0 & -- & 0.904 \\
ERANN-1-6 & $\times$& 54.1 &--& 0.906\\
ERANN-2-6  &$\times$& 54.5 & --& \textbf{0.908}\\
 \hline
  \end{tabular}}
  \label{tab:urban}
\end{table}

Comparing ERANNs with $s_m=0$ and $s_m \geq 1$, we observe that the proposed technique is effective and helps improve the performance of systems for UrbanSound8K. Moreover, transfer learning significantly improves the performance of ERANNs. 

Our best fine-tuned system \mbox{ERANN-2-6} achieves an accuracy of 0.908, which is better than the performance of the previous best system~\cite{Guzhov2021ESResNeXtfbspLR} with an accuracy of 0.891. Moreover, \mbox{ERANN-2-4} with an accuracy of 0.897 outperforms the previous state-of-the-art system and is 1.24x smaller.

The comparison of the two approaches of CNN-based systems is the same as for ESC-50.

\subsection{RAVDESS}
The RAVDESS dataset~\cite{Livingstone2018TheRA} includes speech and song recordings of 24 professional actors with eight diverse emotions. We use the speech set that comprises~1440 audio recordings with an average duration of~4 seconds. 

For this dataset, we evaluate ERANNs considering 4-fold cross-validation. Folds are formed, so that the specific actor is in one fold for robust results. As for ESC-50, all our models are trained for the same number of iterations, and we repeat all the experiments three times and report the average value.

\subsubsection{Results}

For RAVDESS, we use modified mixup on the waveform, the temporal cropping ($t_c = 3$), and SpecAugment during training. We also use pitch shifting. 

Table~\ref{tab:tablerav} shows the comparison of our ERANNs with various values of hyperparameters and with previous APR systems for RAVDESS. Moreover, the table compares ERANNs trained with and without pitch shifting (``P-S'' in the table).

\begin{table}[th]
  \caption{Comparison of APR systems for RAVDESS}
  \centering
   \scalebox{1.00}{
  \begin{tabular}{ lccccc}
\hline
& \textbf{P-S}  & \textbf{e2e}  & \textbf{Params} &  \multicolumn{2}{c}{\textbf{Accuracy}} \\
    
& & &\textbf{\footnotesize{$\bf{\times \, 10^6}$}} &\textbf{\footnotesize{Scratch}} & \textbf{\footnotesize{Fine-tune}} \\ \hline

CNN14~\cite{9229505} & -- & $\times$ & 79.7 & 0.692 &  0.721 \\ 
 \hline
ERANN-1-3 & $\times$ & $\times$ & 13.5 & 0.669 & --\\
ERANN-1-3 & $\checkmark$ & $\times$ & 13.5 & 0.731 & --\\
ERANN-0-4 & $\times$ & $\times$ & 24.0 & 0.705 & -- \\
ERANN-0-4 & $\checkmark$ & $\times$& 24.0 &  \textbf{0.748} & -- \\
ERANN-1-4 &$\checkmark$ & $\times$ & 24.1 &0.741 & --\\
ERANN-2-4 & $\checkmark$ & $\times$ & 24.2 & 0.722 & 0.730 \\
ERANN-2-5 &$ \checkmark$ & $\times$ & 37.9 & 0.720 & 0.737\\
ERANN-0-6 &$\checkmark$ & $\times$ & 54.0& -- & \textbf{0.743}\\
ERANN-1-6 & $\checkmark$ & $\times$ &  54.1 &--& 0.735\\
ERANN-2-6  & $\checkmark$ & $\times$ & 54.5 & --& 0.734\\

 \hline
  \end{tabular}}
  \label{tab:tablerav}
\end{table}

As listed in Table~\ref{tab:tablerav}, using pitch shifting during training significantly improves the performance of systems. Pre-training on the AudioSet dataset does not improve the performance of systems on RAVDESS. Moreover, our ERANNs with $s_m > 0$ have worse performance than ERANNs with $s_m = 0$.

We can conclude that the proposed method helps improve the performance of our systems but not for all APR tasks. Nevertheless, our \mbox{ERANN-0-4} with an accuracy of~0.748 outperforms the previous state-of-the-art system~\cite{9229505} with an accuracy of~0.721 by a large margin. Moreover, our \mbox{ERANN-0-4} is 3.32x smaller than the previous state-of-the-art system.

\subsection{Discussion}
In this article, we have proposed a method for decreasing the computational complexity of CNN-based systems for APR tasks. The proposed method has helped reduce the computational complexity and improve the performance of ERANNs on three audio datasets (AudioSet, ESC-50, and UrbanSound8K). Moreover, this technique can be applied in other CNN architectures to improve their efficiency for APR tasks.

Furthermore, we have shown the impact of various data augmentation techniques and transfer learning on the performance of APR systems. For instance, we have demonstrated the advantage of using modified mixup for AudioSet; however, in previous studies, only standard mixup was used for this dataset.

Owing to the presence of two hyperparameters in our architecture, ERANNs are flexible, and the system with specific values of $W$ and $s_m$ can be selected considering the best trade-off between the computational complexity and the performance on specific APR tasks with the limitation of computational resources. Another advantage of the proposed systems is that ERANNs do not depend on the duration of input audio signals. Moreover, ERANNs achieve high mean average precision on AudioSet and state-of-the-art accuracies on ESC-50, UrbanSound8K, and RAVDESS. Furthermore, the proposed APR systems have less computational complexity than previous state-of-the-art systems for the aforementioned audio datasets.

The main shortcomings of our APR systems are as follows. First, ERANNs do not outperform the best system AST~\cite{gong21b_interspeech} on \hbox{AudioSet}, and second, ERANNs in this study have significantly more computational complexity than the modern end-to-end APR system AemNet-DW~\cite{9414229}. Moreover, we have not considered sound event detection~\cite{Turpault2019_DCASE}, which are, in addition to tagging and classification tasks, an important subtask of APR.

It is also worth mentioning that the reported results were obtained on validation sets and not on independent test sets (as in most previous works), and therefore, they are likely to be biased.

\section{Conclusion}
\label{section5}
We have proposed a new convolutional neural network for audio pattern recognition tasks and a simple technique for decreasing the computational complexity of CNN-based systems. Moreover, we have demonstrated the impact of various data augmentation techniques and transfer learning on the performance of APR systems. Our systems have high results on the AudioSet dataset and new state-of-the-art results on the \mbox{ESC-50}, UrbanSound8K, and RAVDESS datasets. The proposed systems are more efficient than previous best systems.

\section{Acknowledgements}
This work was partly supported by NVIDIA Inception Program for AI Startups; FASIE grant 0060584; RFBR grant 19-29-01175. We also thank the anonymous reviewers from Pattern Recognition Letters, whose suggestions helped improve the manuscript quality.

\bibliographystyle{IEEEtran}
\bibliography{refs}

\begin{thebibliography}{10}
\providecommand{\url}[1]{#1}
\def\UrlFont{\rmfamily}
\providecommand{\newblock}{\relax}
\providecommand{\bibinfo}[2]{#2}
\providecommand\BIBentrySTDinterwordspacing{\spaceskip=0pt\relax}
\providecommand\BIBentryALTinterwordstretchfactor{4}
\providecommand\BIBentryALTinterwordspacing{\spaceskip=\fontdimen2\font plus
\BIBentryALTinterwordstretchfactor\fontdimen3\font minus
  \fontdimen4\font\relax}
\providecommand\BIBforeignlanguage[2]{{%
\expandafter\ifx\csname l@#1\endcsname\relax
\typeout{** WARNING: IEEEtran.bst: No hyphenation pattern has been}%
\typeout{** loaded for the language `#1'. Using the pattern for}%
\typeout{** the default language instead.}%
\else
\language=\csname l@#1\endcsname
\fi
#2}}

\bibitem{10.1145/2733373.2806390}
K.~J. {Piczak}, ``{ESC: Dataset for Environmental Sound Classification},'' in
  \emph{Proceedings of the 23rd ACM International Conference on Multimedia},
  2015, pp. 1015--1018.

\bibitem{10.1145/2647868.2655045}
J.~{Salamon}, C.~{Jacoby}, and J.~P. {Bello}, ``{A Dataset and Taxonomy for
  Urban Sound Research},'' in \emph{Proceedings of the 22nd ACM International
  Conference on Multimedia}, 2014, p. 1041–1044.

\bibitem{Turpault2019_DCASE}
N.~{Turpault}, R.~{Serizel}, A.~{Shah}, and J.~{Salamon}, ``{Sound event
  detection in domestic environments with weakly labeled data and soundscape
  synthesis},'' in \emph{Workshop on Detection and Classification of Acoustic
  Scenes and Events}, 2019.

\bibitem{7952261}
J.~F. {Gemmeke}, D.~P.~W. {Ellis}, D.~{Freedman}, A.~{Jansen}, W.~{Lawrence},
  R.~C. {Moore}, M.~{Plakal}, and M.~{Ritter}, ``{Audio Set: An ontology and
  human-labeled dataset for audio events},'' in \emph{2017 IEEE International
  Conference on Acoustics, Speech and Signal Processing (ICASSP)}, 2017, p.
  776–780.

\bibitem{cite6}
M.~{Vacher}, J.-F. {Serignat}, and S.~{Chaillol}, ``{Sound classification in a
  smart room environment: an approach using GMM and HMM methods},'' in
  \emph{Proceedings of the IEEE Conference on Speech Technology and
  Human-Computer Dialogue}, 2007, p. 135–146.

\bibitem{5202537}
L.~{Ballan}, A.~{Bazzica}, M.~{Bertini}, A.~{Del Bimbo}, and G.~{Serra},
  ``{Deep networks for audio event classification in soccer videos},'' in
  \emph{Proceedings of the IEEE International Conference on Multimedia and
  Expo}, 2009, p. 474–477.

\bibitem{1021072}
G.~{Tzanetakis} and P.~{Cook}, ``{Musical Genre Classification of Audio
  Signals},'' \emph{IEEE Transactions on Speech and Audio Processing}, vol.~10,
  pp. 293--302, 2002.

\bibitem{Livingstone2018TheRA}
S.~{Livingstone} and F.~{Russo}, ``{The Ryerson Audio-Visual Database of
  Emotional Speech and Song ({RAVDESS}): A dynamic, multimodal set of facial
  and vocal expressions in North American English},'' \emph{PLoS ONE}, vol.~13,
  2018.

\bibitem{10.1007/978-981-10-7419-6_6}
e.~a. B.~{Rocha}, ``{A Respiratory Sound Database for the Development of
  Automated Classification},'' in \emph{Precision Medicine Powered by pHealth
  and Connected Health}, 2017, pp. 33--37.

\bibitem{gong21b_interspeech}
Y.~{Gong}, Y.-A. {Chung}, and J.~{Glass}, ``{AST: Audio Spectrogram
  Transformer},'' in \emph{INTERSPEECH}, 2021, pp. 571--575.

\bibitem{9229505}
Q.~{Kong}, Y.~{Cao}, T.~{Iqbal}, Y.~{Wang}, W.~{Wang}, and M.~D. {Plumbley},
  ``{PANNs}: {Large-Scale Pretrained Audio Neural Networks for Audio Pattern
  Recognition},'' \emph{IEEE/ACM Transactions on Audio, Speech, and Language
  Processing}, vol.~28, pp. 2880--2894, 2020.

\bibitem{9414229}
P.~{Lopez-Meyer}, J.~A. {del Hoyo Ontiveros}, H.~{Lu}, and G.~{Stemmer},
  ``{Efficient End-to-End Audio Embeddings Generation for Audio Classification
  on Target Applications},'' in \emph{ICASSP 2021 - 2021 IEEE International
  Conference on Acoustics, Speech and Signal Processing (ICASSP)}, 2021, pp.
  601--605.

\bibitem{Guzhov2021ESResNeXtfbspLR}
A.~{Guzhov}, F.~{Raue}, J.~{Hees}, and A.~R. {Dengel}, ``{ESResNe(X)t-fbsp:
  Learning Robust Time-Frequency Transformation of Audio},'' in \emph{2021
  International Joint Conference on Neural Networks (IJCNN)}, 2021, pp. 1--8.

\bibitem{Pham_2021_CVPR}
H.~{Pham}, Z.~{Dai}, Q.~{Xie}, and Q.~V. {Le}, ``{Meta Pseudo Labels},'' in
  \emph{Proceedings of the IEEE/CVF Conference on Computer Vision and Pattern
  Recognition (CVPR)}, June 2021, pp. 11\,557--11\,568.

\bibitem{5206848}
J.~{Deng}, W.~{Dong}, R.~{Socher}, L.-J. {Li}, K.~{Li}, and L.~{Fei-Fei},
  ``{ImageNet: A large-scale hierarchical image database},'' in \emph{2009 IEEE
  Conference on Computer Vision and Pattern Recognition}, 2009, pp. 248--255.

\bibitem{pmlr-v97-tan19a}
M.~{Tan} and Q.~{Le}, ``{Efficient{N}et: Rethinking Model Scaling for
  Convolutional Neural Networks},'' in \emph{The 36th International Conference
  on Machine Learning (ICML)}, vol.~97, 2019, pp. 6105--6114.

\bibitem{zagoruyko2017wide}
S.~{Zagoruyko} and N.~{Komodakis}, ``{Wide Residual Networks},'' 2017.

\bibitem{Park2019SpecAugmentAS}
W.~C. Daniel S.~{Park}, Y.~{Zhang}, C.-C. {Chiu}, B.~{Zoph}, E.~D. {Cubuk}, and
  Q.~V. {Le}, ``{SpecAugment: A Simple Data Augmentation Method for Automatic
  Speech Recognition},'' in \emph{INTERSPEECH}, 2019.

\bibitem{tokozume2018learning}
Y.~{Tokozume}, Y.~{Ushiku}, and T.~{Harada}, ``{Learning from Between-class
  Examples for Deep Sound Recognition},'' in \emph{International Conference on
  Learning Representations}, 2018.

\bibitem{810857}
J.~{Laroche} and M.~{Dolson}, ``{New phase-vocoder techniques for
  pitch-shifting, harmonizing and other exotic effects},'' in \emph{Proceedings
  of the 1999 IEEE Workshop on Applications of Signal Processing to Audio and
  Acoustics. WASPAA'99 (Cat. No.99TH8452)}, 1999, pp. 91--94.

\bibitem{Vuegen2013ANMA}
L.~{Vuegen}, B.~{Broeck}, P.~{Karsmakers}, J.~F. {Gemmeke}, B.~{Vanrumste}, and
  H.~V. {Hamme}, ``{An MFCC-GMM approach for event detection and
  classification},'' in \emph{IEEE Workshop on Applications of Signal
  Processing to Audio and Acoustics (WASPAA)}, 2013.

\bibitem{9413035}
A.~{Guzhov}, F.~{Raue}, J.~{Hees}, and A.~{Dengel}, ``{ESResNet: Environmental
  Sound Classification Based on Visual Domain Models},'' in \emph{2020 25th
  International Conference on Pattern Recognition (ICPR)}, 2021, pp.
  4933--4940.

\bibitem{rnn_comparison}
J.~{Li}, W.~{Dai}, F.~{Metze}, S.~{Qu}, and S.~{Das}, ``{A comparison of Deep
  Learning methods for environmental sound detection},'' in \emph{2017 IEEE
  International Conference on Acoustics, Speech and Signal Processing
  (ICASSP)}, 2017, pp. 126--130.

\bibitem{ABDOLI2019252}
S.~Abdoli, P.~Cardinal, and A.~{Lameiras Koerich}, ``{End-to-end environmental
  sound classification using a 1D convolutional neural network},'' \emph{Expert
  Systems with Applications}, vol. 136, pp. 252--263, 2019.

\bibitem{7780459}
K.~{He}, X.~{Zhang}, S.~{Ren}, and J.~{Sun}, ``{Deep Residual Learning for
  Image Recognition},'' \emph{2016 IEEE Conference on Computer Vision and
  Pattern Recognition (CVPR)}, pp. 770--778, 2016.

\bibitem{10.1007/978-3-319-46493-0_38}
K.~{He}, X.~{Zhang}, S.~{Ren}, and J.~{{Sun}}, ``{Identity Mappings in Deep
  Residual Networks},'' in \emph{Computer Vision -- ECCV 2016}, 2016, pp.
  630--645.

\bibitem{Lu_2020}
L.~{Lu}, Y.~{Shin}, Y.~{Su}, and G.~{Karniadakis}, ``{Dying ReLU and
  Initialization: Theory and Numerical Examples},'' \emph{Communications in
  Computational Physics}, vol.~28, p. 1671–1706, 2020.

\bibitem{DBLP:journals/corr/KingmaB14}
D.~P. {Kingma} and J.~{Ba}, ``{Adam: {A} Method for Stochastic Optimization},''
  in \emph{3rd International Conference on Learning Representations, {ICLR}
  2015, San Diego, CA, USA, May 7-9, 2015, Conference Track Proceedings}, 2015.

\bibitem{smith2018superconvergence}
L.~N. {Smith} and N.~{Topin}, ``{Super-Convergence: Very Fast Training of
  Neural Networks Using Large Learning Rates},'' in \emph{Artificial
  Intelligence and Machine Learning for Multi-Domain Operations Applications},
  vol. 11006, 2019, pp. 369--386.

\end{thebibliography}

\end{sloppy}
\end{document}